\documentclass[10pt,letterpaper]{article}
\usepackage[top=0.85in,left=2.75in,footskip=0.75in]{geometry}

\usepackage{amsmath,amssymb}

\usepackage{changepage}

\usepackage{textcomp,marvosym}

\usepackage{cite}

\usepackage{nameref,hyperref}

\usepackage[right]{lineno}

\usepackage[nopatch=eqnum]{microtype}
\DisableLigatures[f]{encoding = *, family = * }

\usepackage[table]{xcolor}

\usepackage{array}

\newcolumntype{+}{!{\vrule width 2pt}}

\newlength\savedwidth

\raggedright
\usepackage[aboveskip=1pt,labelfont=bf,labelsep=period,justification=raggedright,singlelinecheck=off]{caption}

\makeatletter
\renewcommand{\@biblabel}[1]{\quad#1.}
\makeatother

\usepackage{lastpage,fancyhdr,graphicx}
\usepackage{epstopdf}
\fancyheadoffset[L]{2.25in}
\fancyfootoffset[L]{2.25in}
\begin{document}
\vspace*{0.2in}

\begin{flushleft}
{\Large
\textbf\newline{Global drivers and barriers to the public acceptance of autonomous vehicles: Evidence from 17 countries} 
}
\newline
\\
Antonios Saravanos\textsuperscript{1}
\\
\bigskip
\textbf{1} Division of Applied Undergraduate Studies, New York University, New York, NY, USA
\\
\bigskip

%
%

\textcurrency Current address: 7 East 12th Street, Room 625B, New York University, New York, NY 10003, USA. 

*saravanos@nyu.edu

\end{flushleft}
\section*{Abstract}

This study investigated the public acceptance of Society of Automotive Engineers Level~3 conditionally automated cars, which can self-drive under certain specified conditions but require the human driver to remain ready to resume control when requested. Previous Unified Theory of Acceptance and Use of Technology 2 (UTAUT2)-based research has focused mainly on European samples, and so it is still unclear whether the same factors shape acceptance across broader world regions. This knowledge gap was addressed using the L3Pilot Global User Acceptance Survey. From an original dataset of 18,631 respondents, the final analytic sample comprised 18,603 respondents from 17 countries across Africa, Asia, Europe, North America, and South America. The data were analyzed using a UTAUT2-based structural equation model to examine how performance expectancy, effort expectancy, social influence, facilitating conditions, and hedonic motivation shape the intention to use Level~3 cars. The model showed strong explanatory power. Across the analytic sample, the intention to use Level~3 cars was driven mainly by performance expectancy, social influence, and hedonic motivation. Effort expectancy and facilitating conditions also contributed, but they played smaller direct roles. Age, gender, and previous experience with advanced driver assistance systems were statistically significant, but comparatively weak predictors. Overall, the findings suggest that the acceptance of Level~3 automated cars depends less on demographic characteristics or ease-of-use concerns and more on whether people see the technology as useful, socially supported, and enjoyable to use.

\nolinenumbers

\section*{Introduction}

Conditionally automated vehicles (Society of Automotive Engineers (SAE) Level~3) occupy a distinctive position in the automation landscape. Within defined operational design domains, the automated driving system can perform the dynamic driving task, but the human driver remains the fallback user and must resume control when requested. Because Level~3 driving combines the promise of automation with continuing human responsibility, public acceptance is not a peripheral issue, rather being central to diffusion, safe use, product design, regulation, and communication.

Research on public attitudes toward automated driving has expanded rapidly, but the evidence base remains uneven and difficult to compare across countries, automation levels, and measurement approaches. An early international survey of respondents from 109 countries showed substantial cross-national variation in attitudes toward automated driving, together with persistent concerns about safety, legal liability, and software misuse\cite{Kyriakidis2015}. Recent reviews have similarly concluded that autonomous vehicle (AV)-acceptance research remains methodologically heterogeneous, often being dominated by descriptive or single-context designs\cite{Zhang2023,Kaye2021}. Meta-analytic evidence has identified trust, perceived usefulness, perceived risk, safety perceptions, enjoyment, social influence (SI), and perceived ease of use as recurring determinants of acceptance, while also showing that the strength of these relationships varies by region, automation level, age group, ownership model, and cultural context\cite{Zhang2021,Gopinath2022}. This literature indicates that acceptance is shaped by multiple psychological mechanisms, and that harmonized multi-country evidence is especially valuable.

The need for more focused evidence is particularly important for SAE Level~3 passenger cars. Unlike fully driverless systems, Level~3 vehicles retain a supervisory role for the driver, users being able to disengage from parts of the driving task during active automation, while still being aware of the system boundaries and being ready to respond appropriately to take-over requests. This has created distinctive acceptance challenges related to trust calibration, perceived safety, controllability, human--machine interaction, and take-over readiness. Evidence from a field experiment showed that trust and both perceived usefulness and safety predicted acceptance after exposure to a Level~3 vehicle, with perceived ease of use becoming more important once the users had directly experienced the technology\cite{Xu2018}. A recent review of trust in Level~3 automation further highlighted the role of system reliability, transparency, human--machine-interface design, environmental complexity, non-driving tasks, and take-over performance in shaping user confidence\cite{Alkurdi2025}. The acceptance of Level~3 cars should therefore be treated as a distinct phenomenon rather than information being inferred from studies that aggregated Levels~3--5.

Several behavioral theories can be applied in the study of Level~3 vehicle acceptance, including the technology acceptance model, theory of planned behavior, diffusion of innovations theory, innovation resistance theory, and behavioral reasoning theory\cite{Davis1989,Ajzen1991,Rogers2003,Arpaci2024,Arpaci2025}. These frameworks highlight useful mechanisms, such as perceived usefulness, perceived ease of use, attitudes, subjective norms, perceived behavioral control, innovation attributes, resistance factors, and reasons for or against adoption. For this work, however, the Unified Theory of Acceptance and Use of Technology 2 (UTAUT2) provided a particularly suitable framework because it combines several technology-specific acceptance mechanisms in one parsimonious model. The UTAUT2 links behavioral intention (BI) to performance expectancy (PE), effort expectancy (EE), SI, facilitating conditions (FC), hedonic motivation (HM), and user characteristics, such as age, gender, and experience\cite{Venkatesh2012}. In the context of Level~3 automation, these constructs map directly onto perceived functional benefits, ease of learning and use, social endorsement, perceived support and resources, anticipated enjoyment, and heterogeneous user backgrounds. Previous UTAUT2-based research on conditionally automated cars has shown that HM, SI, and PE were important predictors in a large European driver sample\cite{Nordhoff2020}, and more recent multi-country evidence has again highlighted PE, trust, and SI as central determinants of Level~3 acceptance\cite{Nordhoff2026}.

Recent work has also broadened AV acceptance research by connecting automated vehicles with sustainability-oriented adoption, green behavior, and connected-vehicle ecosystems. Arpaci et al.\cite{Arpaci2026} modeled sustainable AV adoption intentions by extending the theory of planned behavior to include environmental concerns, authority support, and economic benefits. A related study combined innovation resistance theory, partial least squares structural equation modeling (PLS-SEM), and artificial neural networks to examine green behavior and environmental sustainability in AVs, identifying environmental benefits, environmental concerns, economic benefits, and technophilia as drivers, and cost, security, and privacy concerns as barriers\cite{Arpaci2024}. Extending the discussion to connected mobility, Arpaci et al.\cite{Arpaci2025} used behavioral reasoning theory to examine vehicle-to-everything (V2X) adoption, finding that environmental values shaped reasons both for and against adoption, which in turn influenced attitudes, BI, and green behavior. These studies complemented the UTAUT2 by showing that AV acceptance is not only a matter of usefulness, ease of use, SI, and enjoyment, but also of environmental values, economic value, institutional support, security and privacy concerns, and technology-specific reasons for and against adoption.

Despite these advances, important knowledge gaps remain. Previous studies have made valuable contributions by examining international public opinion\cite{Kyriakidis2015}, synthesizing the broader AV acceptance literature\cite{Zhang2021,Zhang2023}, modeling Level~3 acceptance in European samples\cite{Nordhoff2020}, analyzing intentions to use Level~3 functions across driving environments and countries\cite{Louw2021}, and identifying broad user profiles, such as enthusiastic, neutral, and skeptical respondents\cite{Nordhoff2022}. However, much of the evidence is based on single countries, regional samples, descriptive cross-country comparisons, specific use cases, or segmentation analyses. The field therefore still lacks a pooled structural analysis that uses the full first-phase L3Pilot multinational dataset to quantify the relative importance of the main UTAUT2 acceptance constructs for Level~3 cars across a broad international sample\cite{L3PilotD7}.

This gap was addressed by estimating a UTAUT2-based PLS-SEM model using harmonized responses from 18{,}631 licensed drivers in 17 countries across Africa, Asia, Europe, North America, and South America. The novelty of the study is threefold. First, unlike previous European-focused UTAUT2 applications, this study estimated one pooled structural model across the full multinational sample. Second, whereas previous L3Pilot-based works have mainly examined country-level intentions, driving-environment-specific intentions, or acceptance profiles\cite{Louw2021,Nordhoff2022}, this study quantified the direct and indirect effects of PE, EE, SI, FC, and HM in one model. Third, by comparing direct and total effects, this study identified both the strongest immediate predictors of BI and the upstream constructs that shape intention through indirect pathways. The analysis therefore provides a broad pooled estimate of the main acceptance mechanisms for Level~3 cars, rather than a test of country- or region-specific differences.

This study contributes to the literature in three ways. Empirically, it provides one of the broadest harmonized pooled analyses of the public acceptance of SAE Level~3 cars, based on the first-phase L3Pilot Global User Acceptance Survey\cite{L3PilotD7}. Theoretically, it clarifies the relative importance of the core UTAUT2 constructs for conditional automation and situates these constructs within newer work on sustainable AV adoption, green behavior, and V2X-enabled mobility\cite{Arpaci2024,Arpaci2025,Arpaci2026}. Methodologically, it combines pooled PLS-SEM with direct and total effect estimation to distinguish proximal predictors of BI from constructs that operate partly through indirect pathways. At the same time, because the L3Pilot dataset does not directly measure environmental values, perceived economic benefits, authority support, cost barriers, cybersecurity concerns, privacy concerns, or V2X-specific reasons for and against adoption, these factors can be used to position and assess the findings rather than being observed predictors in the empirical model.

\section*{Materials and methods}

In this section, the data source is described, as well as, survey procedure, ethical considerations, data preparation, measures, model specification, and PLS-SEM procedures used in the study. The section is limited to methodological information; item-retention decisions, measurement-model statistics, structural coefficients, moderation results, and predictive-performance findings are reported in the Results section.

\subsection*{Data source and survey procedure}

The publicly available L3Pilot Global User Acceptance Survey, First Phase Data was used. A pooled online survey dataset collected during the first phase of the European Union (EU)-funded L3Pilot project\cite{Lehtonen2021,L3PilotD7,Louw2021}. The dataset combines responses from the first two data-collection waves and focuses on the public acceptance of SAE Level~3 conditionally automated cars\cite{Lehtonen2021,L3PilotD7}. To ensure version-specific reproducibility, the analyses reported here were based on Zenodo version 1.2 of the first-phase dataset (doi: 10.5281/zenodo.8389544), accessed on January 10, 2026.

The original first-phase dataset included 18{,}631 licensed drivers from 17 countries, including the United Kingdom, the United States, Sweden, Germany, France, China, Hungary, Italy, Finland, Spain, Brazil, Indonesia, India, Japan, Russia, South Africa, and Turkey. Data collection was administered primarily by the market research institute INNOFACT AG using the EXAVO survey platform, with the Finnish data being collected by Taloustutkimus Oy using its nationally representative internet panel\cite{L3PilotD7,Louw2021,Nordhoff2020}. Respondents were recruited from existing online access panels through email invitations containing a survey link. Quota-based market-research sampling criteria were used to approximate national distributions of age, gender, and income, and data collection in each country was closed once the relevant quotas had been filled\cite{L3PilotD7,Louw2021}.

Screening questions were used to focus the survey on current and relatively frequent car drivers. Individuals who reported that they ``almost never'' drove a private car, car-sharing vehicle, or rental car were excluded from participating\cite{Nordhoff2020,L3PilotD7,Louw2021}. The survey providers also applied project-level data-quality checks, including measures to prevent bot participation and repeated participation, as well as excluding respondents who selected all transport modes as frequently used, reported atypical travel patterns, answered all knowledge questions on conditionally automated driving with ``I don't know'', or provided logically inconsistent sociodemographic information\cite{Nordhoff2020,L3PilotD7,Louw2021}.

Before launch, the questionnaire was translated into the national or predominant language of each participating country using professional translation procedures. It was then pre-tested in several iterations and soft-launched to address wording and implementation issues\cite{L3PilotD7,Louw2021}. Before answering the attitudinal questions, the respondents were presented with a standardized textual description of a SAE Level~3 conditionally automated car. This description explained that the vehicle could perform steering, acceleration, and braking under specified conditions, such as motorways, traffic jams, urban roads, or parking situations; that the driver could engage in non-driving-related activities while automation was active; and that the system could issue a request to intervene, requiring the driver to resume control\cite{Nordhoff2020,L3PilotD7}.

The questionnaire contained five main parts: (1) sociodemographic characteristics and mobility behavior; (2) knowledge and familiarity with automated driving; (3) attitudes toward Level~3 cars, including UTAUT2 constructs; (4) intended use of Level~3 functions in specific driving environments; and (5) additional sociodemographic and travel-behavior questions\cite{Nordhoff2020,L3PilotD7}. The respondents were informed that the survey was part of the L3Pilot project, that participation was voluntary, and that their responses would be treated anonymously\cite{Nordhoff2022,L3PilotD7}. The respondents generally received small incentives for their participation, these typically being vouchers worth between €0.80 and 1.00 or comparable panel rewards, with the Finnish panel members being offered the chance to win prizes instead\cite{L3PilotD7,Louw2021}.

\subsection*{Ethics statement}

The original L3Pilot Global User Acceptance Survey collected anonymized survey data in accordance with applicable EU and national data-protection requirements, with the respondents providing electronic informed consent prior to participation\cite{L3PilotD7,Nordhoff2020}. With the present study being a secondary analysis of a fully anonymized publicly available dataset, there was no direct contact with the respondents. The New York University Institutional Review Board determined that our analysis did not constitute human-subjects research and therefore required no further Institutional Review Board review (IRB-FY2026-11021).

\subsection*{Data preparation}

The starting dataset included 18{,}631 licensed drivers. Gender was originally coded as 1 = ``Male'', 2 = ``Female'', and 3 = ``Other''. Those respondents coded as 3 = ``Other'' were excluded because this subgroup was very small ($n = 28$) and could not support statistically reliable separate analysis in the present model. The final analytic sample therefore included 18{,}603 respondents coded as 1 = ``Male'' or 2 = ``Female'' on the retained gender variable. This analytic restriction was used for statistical stability and should not be interpreted as implying that gender is inherently binary.

The retained sample was nearly evenly gender balanced, with 9{,}297 respondents coded as male (49.98\%) and 9{,}306 coded as female (50.02\%). The respondents ranged in age from 18 to 69 years ($M = 40.73$, $SD = 13.63$). Previous advanced driver assistance systems (ADAS)-use experience ranged from 0 to 9, with higher values indicating experience with a greater number of ADAS. For the experience measure, omitted responses were treated as indicating no previous experience with the corresponding system. The mean ADAS-use experience score was 1.82 ($SD = 2.48$), with a median of 1.

\begin{figure}[!ht]
\centering
\includegraphics[width=\textwidth]{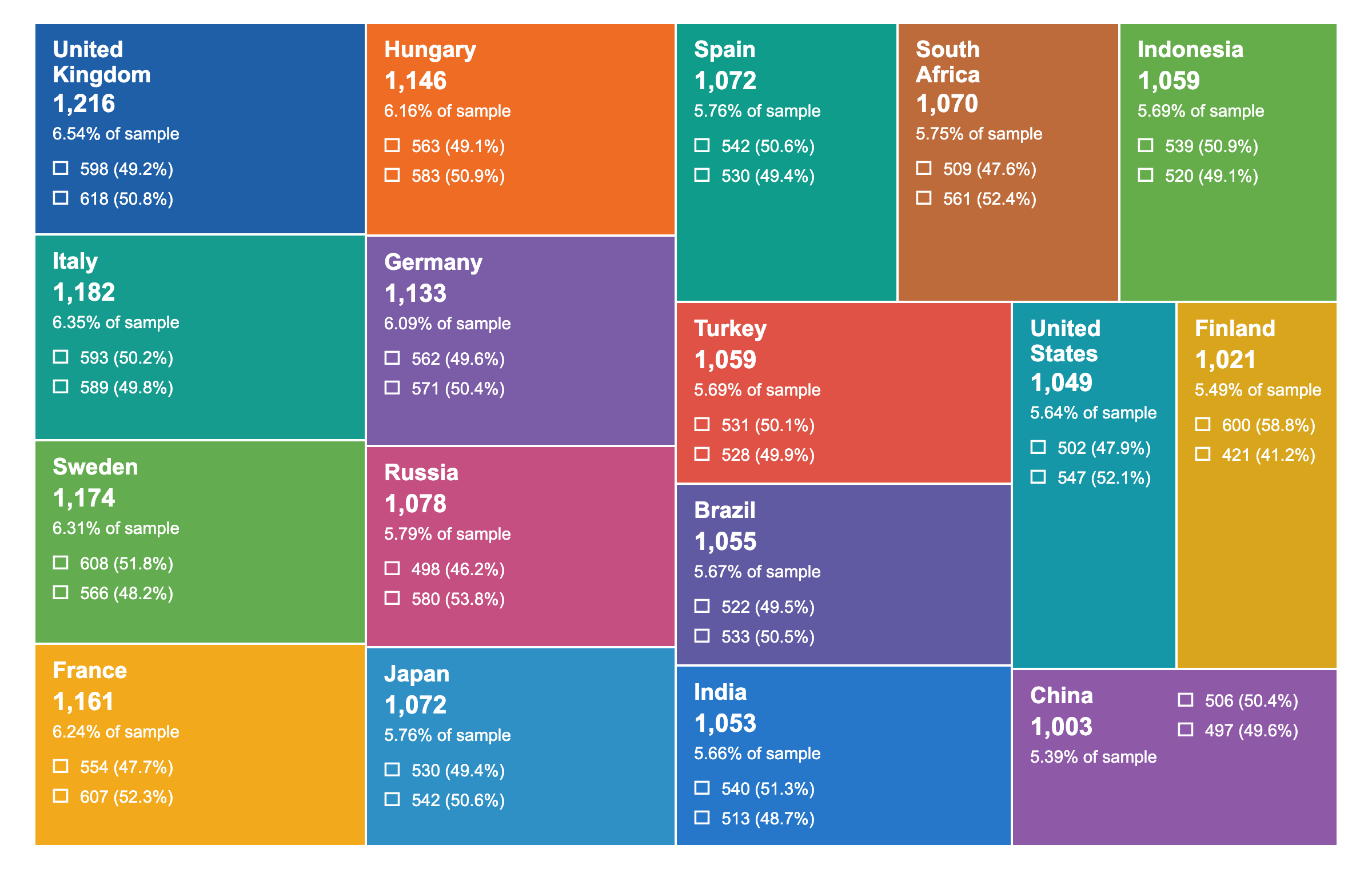}
\caption{\textbf{Distribution of respondents by country.}
The final analytic sample included 18{,}603 respondents from 17 countries. National sample sizes were broadly comparable across countries, ranging from 1{,}003 respondents in China to 1{,}216 respondents in the United Kingdom. Percentages indicate each country's share of the final analytic sample.}
\label{fig:country_distribution}
\end{figure}

\begin{figure}[!ht]
\centering
\includegraphics[width=\textwidth]{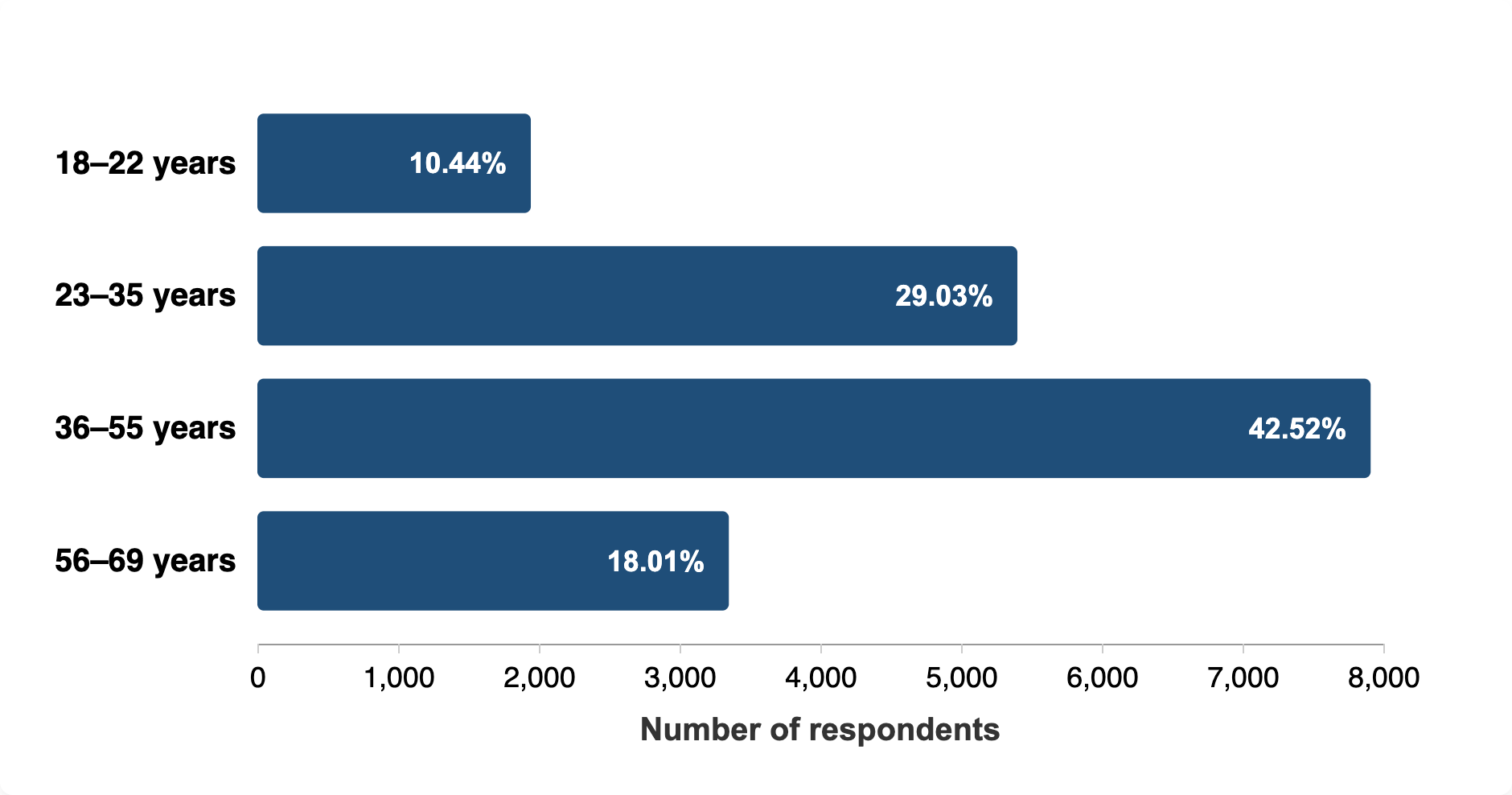}
\caption{\textbf{Distribution of respondents by age.}
The largest age group was 36--55 years, representing 42.52\% of the respondents, followed by 23--35 years at 29.03\%, 56--69 years at 18.01\%, and 18--22 years at 10.44\%.}
\label{fig:age_distribution}
\end{figure}

\begin{figure}[!ht]
\centering
\includegraphics[width=\textwidth]{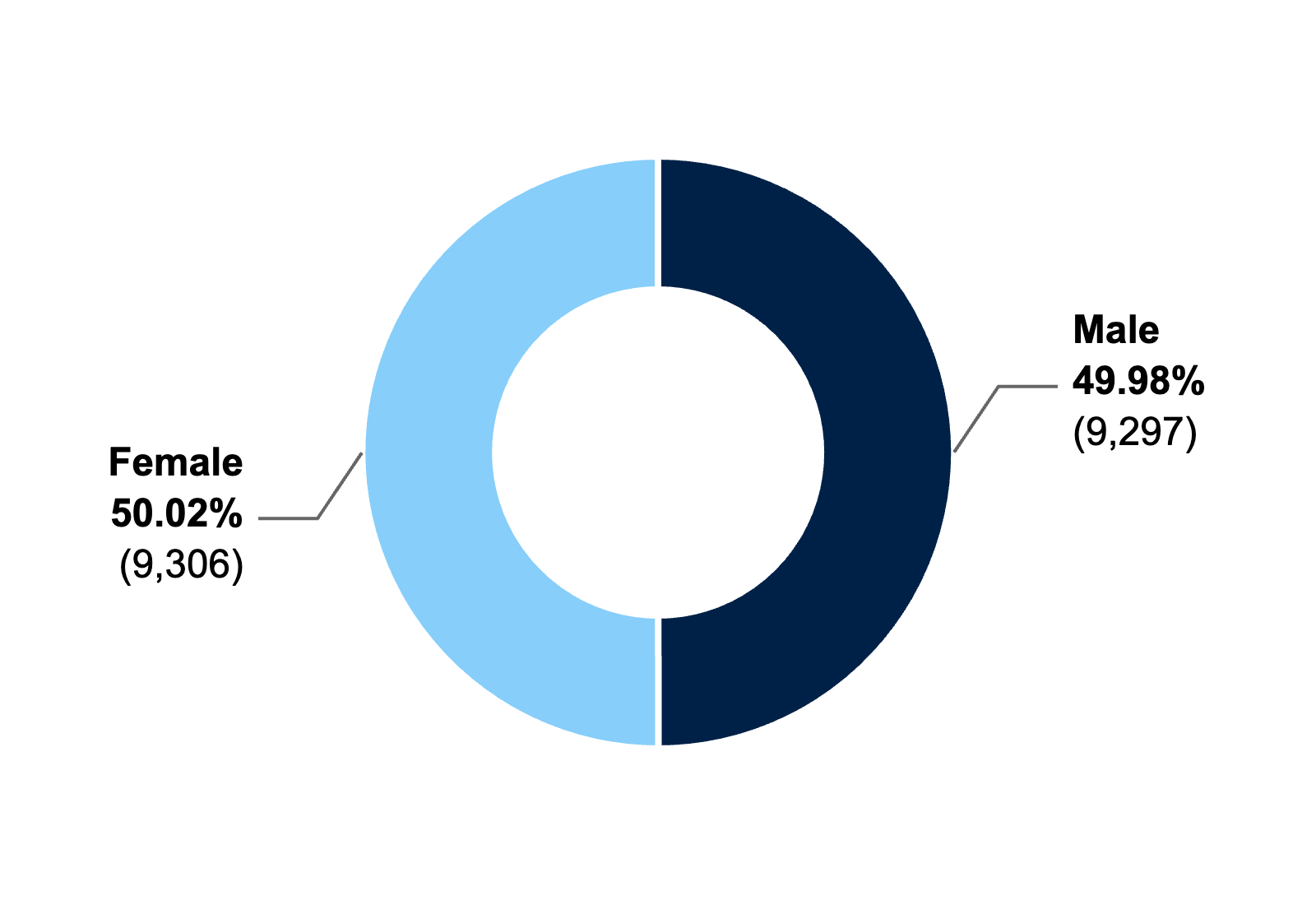}
\caption{\textbf{Distribution of respondents by gender.}
The retained analytic sample was nearly evenly gender balanced, with 9{,}297 respondents coded as male and 9{,}306 respondents coded as female. Respondents coded as ``Other'' were excluded from the analytic sample because the subgroup was very small and could not support statistically reliable separate analysis.}
\label{fig:gender_distribution}
\end{figure}

The demographic composition of the final analytic sample is summarized in Fig~\ref{fig:country_distribution}, Fig~\ref{fig:age_distribution}, and Fig~\ref{fig:gender_distribution}. The figures show the distribution of the respondents across countries, age groups, and gender categories, respectively.

Those responses coded as ``I prefer not to respond'' were treated as missing values for the attitudinal indicators. Following data preparation, the final analytic dataset contained 255 missing values across 1{,}004{,}562 data points (0.03\% overall missingness). The missing data were confined to a single indicator, BI5, with 255 respondents (1.37\%) submitting no response. All remaining variables were complete. Given the extremely low proportion of missing data, the level of missingness was considered negligible and unlikely to bias the results. Accordingly, the remaining missing values were handled in SmartPLS~4 using the software's mean replacement procedure, with missing values being replaced with the corresponding indicator mean prior to model estimation.

\subsection*{Measures}

The measures were taken from the attitudinal and background sections of the L3Pilot Global User Acceptance Survey. The attitudinal part of the questionnaire adapted the UTAUT2 to the context of SAE Level~3 conditionally automated cars, using established acceptance items and previous automated-vehicle research as the basis for the instrument\cite{Venkatesh2012,Xu2018,Nordhoff2020,L3PilotD7}. For this work a new item-to-construct classification was not developed. Instead, the same mapping from L3Pilot questionnaire items to UTAUT2 constructs as proposed by Nordhoff et al.\cite{Nordhoff2020} were used.

Unless otherwise stated, the respondents answered the attitudinal items on five-point agreement scales ranging from 1 = ``strongly disagree'' to 5 = ``strongly agree''. All substantive indicators were coded so that higher values represented stronger agreement with the underlying construct. Responses coded as 6, ``I prefer not to respond'', were not treated as part of the agreement scale and were recoded as missing before model estimation.

The latent measurement model centered on five UTAUT2 antecedents of acceptance and one outcome construct. Performance expectancy captured the extent to which the respondents expected a conditionally automated car to provide practical benefits, such as making travel more useful, comfortable, convenient, or effective. Effort expectancy referred to the anticipated ease of learning, understanding, and using such a vehicle. Social influence captured perceived approval or encouragement from important others. Facilitating conditions reflected whether the respondents believed that the necessary resources, knowledge, assistance, and compatible conditions would be available for use. Hedonic motivation captured the expected enjoyment and positive affect associated with conditional automation. Behavioral intention was specified as the dependent latent construct and represented the respondents' stated intention to use or adopt a conditionally automated car in the future, including future-use and purchase-oriented intentions\cite{Venkatesh2012,Nordhoff2020,Louw2021}.

All latent variables were modeled as reflective constructs because the indicators were treated as observable manifestations of the underlying acceptance beliefs rather than as separate formative components\cite{Hair2019}. The reflective measurement model was evaluated using standard PLS-SEM criteria for indicator reliability, internal consistency reliability, convergent validity, and discriminant validity. The final retained indicators, item-removal decisions, outer loadings, reliability statistics, and validity evidence are reported in the Results and Supporting information sections.

The full UTAUT2 framework also includes price value and habit\cite{Venkatesh2012}. These constructs were not included as formal latent predictors in the present model because the L3Pilot first-phase instrument did not provide comparable multi-item measures for them in this context, and because habitual use is difficult to interpret for respondents without routine direct experience of Level~3 conditional automation\cite{Nordhoff2022,L3PilotD7}. Cost-related information, such as willingness to pay, was therefore treated as auxiliary descriptive information rather than as a latent UTAUT2 construct in the structural model.

The model also included age, gender, and previous ADAS-use experience as observed background variables. Age was self-reported using six ordered categories (1 = ``under 18 years'', 2 = ``18--22 years'', 3 = ``23--35 years'', 4 = ``36--55 years'', 5 = ``56--69 years'', and 6 = ``over 69 years''). In the PLS-SEM analysis, age was entered as a single-indicator observed ordinal variable using this ordered category coding. Higher values therefore indicate membership in an older age category, not a one-year increase in age.

Gender was entered in SmartPLS~4 as a single-indicator observed variable using the retained coding (1 = ``Male'' and 2 = ``Female''). Higher values therefore indicate female respondents under the coding used in the analysis.

Advanced driver assistance systems-use experience was measured as a summative count of the respondents' previous use experience with nine ADAS: (1) automated emergency braking; (2) forward collision warning; (3) blind-spot monitoring; (4) drowsy driver detection; (5) lane departure warning; (6) lane keeping assistance; (7) adaptive cruise control; (8) parking assist; and (9) self-parking assist. For each technology, the respondents received a value of 1 if they reported having experience using the system and 0 otherwise. The nine binary indicators were then summed to create an ADAS-use experience count ranging from 0 to 9, where higher values indicate experience using a greater number of ADAS technologies. This variable was entered in the PLS-SEM model as a single-indicator observed variable rather than as a reflective latent construct.

\subsection*{Model specification and hypotheses}

The UTAUT2-based hypothesis structure used by Nordhoff et al.\cite{Nordhoff2020} was retained. The same hypothesized relationships were applied to the pooled first-phase L3Pilot multinational dataset analyzed here. Thus, the contribution of the present analysis lies in applying the Nordhoff et al.\cite{Nordhoff2020} UTAUT2 model structure to the broader pooled 17-country sample and estimating direct, indirect, total, and moderation effects, rather than proposing a new theoretical model.

\begin{figure}[!ht]
\centering
\includegraphics[width=\textwidth]{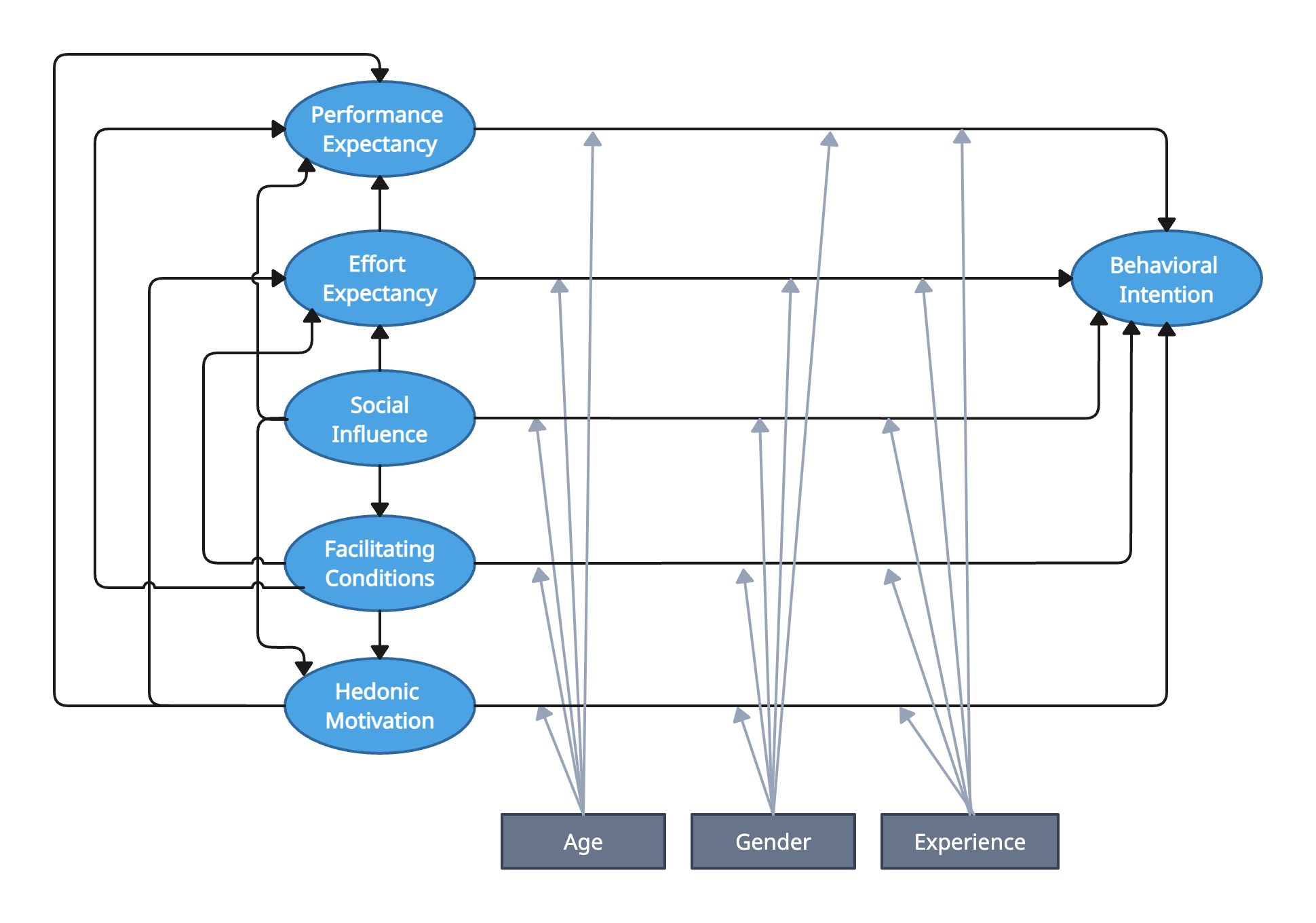}
\caption{\textbf{Conceptual framework of the UTAUT2-based model used in this study.}
The conceptual model illustrates the hypothesized relationships among performance expectancy (PE), effort expectancy (EE), social influence (SI), facilitating conditions (FC), hedonic motivation (HM), and behavioral intention (BI). Age, gender, and previous ADAS-use experience were modeled as moderators of the relationships between the UTAUT2 constructs and BI. The model was adopted from Nordhoff et al.\cite{Nordhoff2020}.}
\label{fig:conceptual_model}
\end{figure}

The structural model included direct paths from EE, FC, HM, PE, and SI to BI. It also included hypothesized interrelations among the UTAUT2 predictor constructs, including paths from EE to PE; FC to EE, HM, and PE; and SI to EE, FC, HM, and PE. Age, gender, and ADAS-use experience were included as observed predictors of BI and were also tested as moderators of the five main UTAUT2 predictor--intention relationships. The conceptual framework adopted in this study is shown in Fig~\ref{fig:conceptual_model}.

The adopted hypotheses are summarized in Table~\ref{tab:adopted_hypotheses}. Because the data are cross-sectional survey responses, statistically supported paths were interpreted as associations consistent with the adopted hypotheses rather than as definitive causal effects.

\begin{table}[!ht]
\centering
\caption{Hypotheses adopted from Nordhoff et al.\cite{Nordhoff2020}}
\label{tab:adopted_hypotheses}

\setlength{\tabcolsep}{4pt}

\begin{tabular}{
>{\raggedright\arraybackslash}p{0.10\textwidth}
>{\raggedright\arraybackslash}p{0.20\textwidth}
>{\raggedright\arraybackslash}p{0.62\textwidth}
}
\hline
\textbf{Hyp.} & \textbf{Path/component} & \textbf{Adopted hypothesis} \\
\hline

\multicolumn{3}{l}{\textit{Direct antecedents of BI}} \\
\hline
H1 & EE $\rightarrow$ BI & EE positively influences BI. \\
H2 & FC $\rightarrow$ BI & FC positively influences BI. \\
H3 & HM $\rightarrow$ BI & HM positively influences BI. \\
H4 & PE $\rightarrow$ BI & PE positively influences BI. \\
H5 & SI $\rightarrow$ BI & SI positively influences BI. \\

\hline
\multicolumn{3}{l}{\textit{Moderating role of user characteristics}} \\
\hline
H6 & AGE, EXPERIENCE, and GENDER as moderators &
AGE, EXPERIENCE, and GENDER moderate the effects of EE, FC, HM, PE, and SI on BI. \\

\hline
\multicolumn{3}{l}{\textit{Interrelations among UTAUT2 constructs}} \\
\hline
H7  & EE $\rightarrow$ PE & EE positively influences PE. \\
H8  & FC $\rightarrow$ PE & FC positively influences PE. \\
H9  & FC $\rightarrow$ EE & FC positively influences EE. \\
H10 & FC $\rightarrow$ HM & FC positively influences HM. \\
H11 & HM $\rightarrow$ PE & HM positively influences PE. \\
H12 & HM $\rightarrow$ EE & HM positively influences EE. \\
H13 & SI $\rightarrow$ PE & SI positively influences PE. \\
H14 & SI $\rightarrow$ EE & SI positively influences EE. \\
H15 & SI $\rightarrow$ FC & SI positively influences FC. \\
H16 & SI $\rightarrow$ HM & SI positively influences HM. \\
\hline
\end{tabular}

\begin{flushleft}
\textit{Notes:} ADAS, advanced driver assistance systems; BI, behavioral intention; EE, effort expectancy; FC, facilitating conditions; HM, hedonic motivation; PE, performance expectancy; SI, social influence. Hypotheses were adopted from the UTAUT2-based L3Pilot model reported by Nordhoff et al.\cite{Nordhoff2020} and applied to the pooled multinational dataset analyzed in the present study.
\end{flushleft}

\end{table}

\subsection*{PLS-SEM analysis}

Variance-based PLS-SEM was used to estimate the relationships between the UTAUT2 constructs, observed background variables, and BI, PLS-SEM being suitable for complex models with correlated predictors and non-normal indicators and able to prioritize prediction-oriented assessment over global model fit\cite{Hair2019}.

All analyses were conducted in SmartPLS~4. Model estimation involved the path-weighting scheme, using two-tailed tests and $\alpha = 0.05$. Bias-corrected bootstrapping with 7{,}000 subsamples was used to obtain standard errors and test statistics for the measurement and structural models.

The reflective measurement model was assessed using indicator reliability, internal consistency reliability, convergent validity, and discriminant validity. Indicator reliability was examined using standardized outer loadings. Internal consistency reliability was evaluated using Cronbach's alpha, composite reliability, and $\rho_A$. Convergent validity was assessed using average variance extracted (AVE), and discriminant validity was examined using the heterotrait--monotrait ratio of correlations (HTMT), cross-loadings, and the Fornell--Larcker criterion\cite{Hair2019,Henseler2015}.

The structural model was evaluated using standardized path coefficients, bootstrapped test statistics, coefficients of determination ($R^2$ and adjusted $R^2$), and total effects on BI. Direct effects were used to identify the strongest proximal predictors of BI, while total effects were used to account for both direct and indirect pathways among the UTAUT2 constructs. Moderation effects were assessed using interaction terms between the observed background variables and the five main UTAUT2 predictors of BI.

Predictive relevance and out-of-sample predictive performance were assessed using PLSpredict, run with 10 folds and 10 repetitions, and the cross-validated predictive ability test (CVPAT) in SmartPLS~4\cite{Ringle2024}. The analysis examined the $Q^2_{\mathrm{predict}}$, root mean square error (RMSE), and mean absolute error (MAE) for the indicators of the endogenous constructs. The PLS-SEM prediction errors were compared with the indicator-average (IA) and linear-model (LM) benchmarks. Positive $Q^2_{\mathrm{predict}}$ values were interpreted as evidence that the PLS-SEM model had outperformed the naive mean-based prediction, whereas lower PLS-SEM RMSE and MAE values were interpreted as evidence of better predictive performance relative to the corresponding benchmark\cite{Shmueli2019}. The CVPAT was additionally used to test whether the average prediction loss of the PLS-SEM model was significantly lower than the IA and LM benchmarks. Negative and statistically significant average loss differences (ALDs) were interpreted as evidence of superior predictive ability relative to a benchmark\cite{Liengaard2021,Sharma2023}.

Overall model fit was assessed descriptively using the SmartPLS~4 model-fit output, including the standardized root mean squared residual (SRMR), $d_{\mathrm{ULS}}$, geodesic distance ($d_G$), chi-squared test, and normed fit index (NFI). Because PLS-SEM is primarily prediction-oriented, these indices were used as supplementary model diagnostics rather than as the main basis for evaluating the model.

\section*{Results}

Using the procedures described above, first the reflective measurement model was evaluated and then the structural model, moderation effects, model fit, and prediction-oriented performance, were assessed.

\subsection*{Measurement model evaluation}

All latent constructs were specified as reflective. The measurement model was evaluated using standard PLS-SEM criteria for indicator reliability, internal consistency reliability, convergent validity, and discriminant validity\cite{Hair2019,Henseler2015}. Overall, the final measurement model met the recommended reliability and convergent-validity criteria after removal of one low-loading PE item. The discriminant validity was more nuanced. The constructs were statistically distinguishable under the HTMT confidence-interval criterion, but several theoretically related construct pairs exceeded the conservative HTMT.90 threshold.

\subsubsection*{Indicator reliability}

Indicator reliability was assessed using standardized outer loadings. The initial measurement model is reported in Table S1, while the final measurement model, after the removal of PE1, is reported in Table S2. Following the common PLS-SEM guidelines, loadings of $\lambda \geq 0.70$ were interpreted as indicating adequate indicator reliability, while items with loadings between 0.40 and 0.70 were considered for removal if doing so improved the reliability or convergent validity without undermining the content validity\cite{Hair2019}.

In the initial specification, PE was measured using five indicators, PE1--PE5. Indicator PE1, which referred to using the time during which a conditionally automated car is driving for other activities, showed the weakest loading among the PE indicators ($\lambda = 0.655$; S1 Table). Because this loading was below the preferred 0.70 threshold, and because the item overlapped conceptually with time-use and hedonic aspects rather than core performance beliefs, PE1 was removed from the final measurement model. This decision is consistent with item-trimming logic in PLS-SEM and with the measurement approach used in the original L3Pilot analysis by Nordhoff et al.\cite{Nordhoff2020}. All subsequent analyses therefore used the final PE specification based on PE2--PE5 only.

After PE1 was removed, all the retained indicators showed satisfactory standardized outer loadings, ranging from 0.729 to 0.905 (S2 Table). The construct-specific loading ranges were 0.729--0.880 for BI, 0.828--0.861 for EE, 0.747--0.820 for FC, 0.889--0.905 for HM, 0.863--0.881 for PE, and 0.826--0.861 for SI. All retained outer loadings were statistically significant at $p < 0.001$.

\subsubsection*{Internal consistency, reliability, and convergent validity}

The construct reliability and convergent validity statistics are summarized in Table~\ref{tab:reliability}. The Cronbach's $\alpha$ values ranged from 0.803 to 0.895, the $\rho_A$ values from 0.805 to 0.895, and the composite reliability values from 0.874 to 0.927. These values exceeded the recommended minimum threshold of 0.70 and remained below 0.95, indicating good internal consistency without suggesting excessive indicator redundancy\cite{Hair2019}. The AVE values ranged from 0.634 to 0.808 and exceeded the recommended cutoff of 0.50 for all reflective constructs. The results therefore support convergent validity for BI, EE, FC, HM, PE, and SI. Age, gender, overall ADAS experience, and the interaction terms involving these variables and the UTAUT2 predictors were specified as single-indicator composites with their outer loadings fixed at 1.00. Consequently, these variables were not evaluated using Cronbach's $\alpha$, composite reliability, or AVE.

\begin{table}[!ht]
\centering
\caption{Construct reliability and convergent validity for reflective constructs.}
\label{tab:reliability}
\begin{tabular}{lcccc}
\hline
\textbf{Construct} & \textbf{Cronbach's $\alpha$} & \boldmath$\rho_A$ & \boldmath$\rho_c$ & \textbf{AVE} \\
\hline
Behavioral Intention (BI)   & 0.881 & 0.888 & 0.914 & 0.680 \\
Effort Expectancy (EE)       & 0.803 & 0.805 & 0.884 & 0.717 \\
Facilitating Conditions (FC) & 0.807 & 0.808 & 0.874 & 0.634 \\
Hedonic Motivation (HM)      & 0.881 & 0.882 & 0.927 & 0.808 \\
Performance Expectancy (PE)  & 0.895 & 0.895 & 0.927 & 0.761 \\
Social Influence (SI)        & 0.868 & 0.872 & 0.910 & 0.717 \\
\hline
\end{tabular}
\begin{flushleft}
\footnotesize \textit{Notes:} $\rho_A$, reliability coefficient; $\rho_c$, composite reliability; AVE, average variance extracted. All constructs met the recommended thresholds (Cronbach's $\alpha$, $\rho_A$, $\rho_c \geq 0.70$; AVE $\geq 0.50$), indicating good internal consistency reliability and convergent validity.
\end{flushleft}
\end{table}

\subsubsection*{Discriminant validity}

Discriminant validity among the reflective UTAUT2 constructs was assessed using the HTMT, bootstrapped 95\% bias-corrected and accelerated confidence intervals based on 7{,}000 subsamples, and indicator cross-loadings\cite{Henseler2015,Hair2019}. Five construct pairs exceeded the conservative HTMT.90 threshold, including PE and BI, SI and BI, EE and FC, HM and BI, and PE and HM (Table~\ref{tab:htmt_ci}). Although these HTMT values indicate substantial overlap among several theoretically related UTAUT2 constructs, the upper bound of each confidence interval remained below 1.00. The constructs are therefore statistically distinguishable according to the HTMT inference criterion, but discriminant validity is only partially supported under the stricter HTMT.90 criterion. This pattern is not unexpected in UTAUT2 applications, where constructs such as PE, BI, SI, and HM may be conceptually and empirically close, particularly in studies of automated vehicle acceptance.

\begin{table}[!ht]
\centering
\caption{HTMT values and bootstrapped 95\% confidence intervals (CIs) for construct pairs with HTMT values above 0.90.}
\label{tab:htmt_ci}
\begin{tabular}{lccc}
\hline
\textbf{Construct pair} & \textbf{HTMT} & \textbf{2.5\% CI} & \textbf{97.5\% CI} \\
\hline
PE $\leftrightarrow$ BI & 0.966 & 0.962 & 0.971 \\
SI $\leftrightarrow$ BI & 0.950 & 0.944 & 0.955 \\
EE $\leftrightarrow$ FC & 0.938 & 0.929 & 0.947 \\
HM $\leftrightarrow$ BI & 0.924 & 0.918 & 0.930 \\
PE $\leftrightarrow$ HM & 0.924 & 0.917 & 0.930 \\
\hline
\end{tabular}

\begin{flushleft}
\footnotesize
\textit{Notes}: HTMT, heterotrait--monotrait ratio of correlations. CIs are bias-corrected bootstrap intervals based on 7{,}000 subsamples. For the HTMT inference criterion against the value of 1.00, discriminant validity is supported when the confidence interval does not include 1.00. Although the upper bounds are below 1.00, all reported pairs exceed the conservative HTMT.90 threshold, indicating substantial construct overlap and suggesting only partial support for discriminant validity.
\end{flushleft}
\end{table}

As an additional indicator-level check, the cross-loadings of all the reflective indicators were inspected (S3 Table). Each retained indicator loaded highest on its intended construct. The smallest margin between an indicator's primary loading and its highest cross-loading was observed for SI4, which loaded 0.826 on SI and 0.768 on BI. Overall, the cross-loading results support indicator-level discriminant validity, while the HTMT results indicate substantial construct overlap for a subset of highly related UTAUT2 dimensions.

\subsection*{Overall model fit}

The model-fit results are reported in Table~\ref{tab:model_fit}. For the estimated model, the SRMR was 0.048---below the commonly used conservative threshold of 0.08 and indicating good approximate model fit. The saturated-model SRMR was similarly low at 0.047. The estimated-model NFI was 0.886---slightly below the commonly cited 0.90 guideline. The $d_{\mathrm{ULS}}$ and $d_G$ values were interpreted cautiously because discrepancy measures are mainly informative when assessed using bootstrap-based exact-fit inference. Overall, the fit indices provide acceptable supplementary diagnostics, while the main evaluation rests on the measurement-model, structural-model, and prediction-oriented criteria reported below.

\begin{table}[!ht]
\centering
\caption{Overall model fit indices.}
\label{tab:model_fit}
\begin{tabular}{lcc}
\hline
\textbf{Fit index} & \textbf{Saturated model} & \textbf{Estimated model} \\
\hline
SRMR               & 0.047     & 0.048 \\
$d_{\mathrm{ULS}}$ & 0.764     & 0.805 \\
$d_G$              & 0.341     & 0.347 \\
Chi-square         & 36001.162 & 36541.086 \\
NFI                & 0.887     & 0.886 \\
\hline
\end{tabular}
\begin{flushleft}
\footnotesize
\textit{Notes}: $d_G$ = geodesic distance; $d_{\mathrm{ULS}}$, squared Euclidean distance; NFI = normed fit index; SRMR, standardized root mean square residual. Fit indices are reported from the SmartPLS~4 model-fit output. In line with the prediction-oriented nature of PLS-SEM, these indices were treated as supplementary diagnostics rather than as the primary basis for model evaluation.
\end{flushleft}
\end{table}

\subsection*{Structural model}

The structural model included PE, EE, SI, FC, and HM as predictors of BI, as in Nordhoff et al.\cite{Nordhoff2020}, with additional paths among the predictor constructs motivated by previous UTAUT2 extensions\cite{Nordhoff2020,L3PilotD7}. The standardized path coefficients, explanatory power using $R^2$ and adjusted $R^2$, and total effects on BI were examined to account for both direct and indirect pathways. Table~\ref{tab:r2} reports the coefficients of determination, Table~\ref{tab:effects} the standardized direct effects and bootstrap test statistics, and Table~\ref{tab:total_effects_bi} summarizes the total effects on BI.

\begin{table}[!ht]
\centering
\caption{Coefficient of determination ($R^2$) for endogenous constructs.}
\label{tab:r2}
\begin{tabular}{lcc}
\hline
\textbf{Construct} & \boldmath$R^2$ & {\boldmath\textbf{Adjusted $R^2$}} \\
\hline
BI & 0.827 & 0.827 \\
EE & 0.616 & 0.616 \\
FC & 0.427 & 0.427 \\
HM & 0.650 & 0.650 \\
PE & 0.770 & 0.770 \\
\hline
\end{tabular}
\end{table}

\subsubsection*{Explained variance}

The structural model showed high explanatory power. The predictors jointly accounted for 82.7\% of the variance in BI ($R^2 = 0.827$), 77.0\% in PE ($R^2 = 0.770$), 65.0\% in HM ($R^2 = 0.650$), 61.6\% in EE ($R^2 = 0.616$), and 42.7\% in FC ($R^2 = 0.427$). Thus, the model explains a substantial share of variance in BI and PE, and a moderate to high share of variance in the remaining endogenous UTAUT2 constructs.

\begin{table}[!ht]
\centering
\caption{Structural model direct effects.}
\label{tab:effects}
\begin{tabular}{lcc}
\hline
\textbf{Path} & $\boldsymbol{\beta}_{\mathrm{direct}}$ & \textbf{\boldmath$t$} \\
\hline
AGE $\rightarrow$ BI & –0.023$^{*}$ & 7.099 \\

EE $\rightarrow$ BI & 0.054$^{*}$ & 4.749 \\
EE $\rightarrow$ PE & 0.112$^{*}$ & 14.854 \\

EXPERIENCE $\rightarrow$ BI & 0.028$^{*}$ & 4.410 \\

FC $\rightarrow$ BI & 0.028$^{*}$ & 2.255 \\
FC $\rightarrow$ EE & 0.538$^{*}$ & 63.415 \\
FC $\rightarrow$ HM & 0.364$^{*}$ & 47.573 \\
FC $\rightarrow$ PE & 0.178$^{*}$ & 21.722 \\

GENDER $\rightarrow$ BI & –0.030$^{*}$ & 4.914 \\

HM $\rightarrow$ BI & 0.185$^{*}$ & 13.214 \\
HM $\rightarrow$ EE & 0.194$^{*}$ & 19.625 \\
HM $\rightarrow$ PE & 0.389$^{*}$ & 45.302 \\

PE $\rightarrow$ BI & 0.382$^{*}$ & 25.615 \\

SI $\rightarrow$ BI & 0.331$^{*}$ & 26.463 \\
SI $\rightarrow$ EE & 0.127$^{*}$ & 15.203 \\
SI $\rightarrow$ FC & 0.653$^{*}$ & 128.554 \\
SI $\rightarrow$ HM & 0.519$^{*}$ & 70.089 \\
SI $\rightarrow$ PE & 0.307$^{*}$ & 41.463 \\
\hline
\end{tabular}

\begin{flushleft}
\footnotesize \textit{Notes}: $\beta_{\mathrm{direct}}$ represents the standardized direct effect for each structural path. Asterisks indicate statistically significant paths. All reported paths are significant at $p < 0.001$ except for FC $\rightarrow$ BI, which is significant at $p < 0.05$.
\end{flushleft}
\end{table}

\subsubsection*{Main predictors of BI}

All five UTAUT2 constructs had positive direct effects on BI (Table~\ref{tab:effects}). The strongest direct predictor was PE ($\beta = 0.382$, $t = 25.615$), followed by SI ($\beta = 0.331$, $t = 26.463$) and HM ($\beta = 0.185$, $t = 13.214$). Effort expectancy showed a smaller positive effect ($\beta = 0.054$, $t = 4.749$), with FC having the weakest positive direct effect ($\beta = 0.028$, $t = 2.255$). These results indicate that intention to use conditionally automated cars was driven mainly by perceived usefulness, social endorsement, and anticipated enjoyment, with ease of use and FC playing smaller direct roles.

The background variables were statistically significant, but comparatively weak, predictors of BI. Age had a small negative effect ($\beta = –0.023$, $t = 7.099$), gender had a small negative effect under the coding used in the analysis ($\beta = –0.030$, $t = 4.914$), and ADAS experience had a small positive effect ($\beta = 0.028$, $t = 4.410$). These coefficients were small compared with the main UTAUT2 predictors, suggesting that demographic and experience-related differences exist, but these are not the primary drivers of BI.

\subsubsection*{Direct effects among UTAUT2 constructs}

The structural model also included positive and statistically significant paths among the UTAUT2 predictor constructs (Table~\ref{tab:effects}). Social influence emerged as the main upstream driver, having a strong positive effect on FC ($\beta = 0.653$, $t = 128.554$), HM ($\beta = 0.519$, $t = 70.089$), and PE ($\beta = 0.307$, $t = 41.463$), as well as a smaller positive effect on EE ($\beta = 0.127$, $t = 15.203$).

Facilitating conditions also played an important upstream role, strongly predicting EE ($\beta = 0.538$, $t = 63.415$) and having additional positive effects on HM ($\beta = 0.364$, $t = 47.573$) and PE ($\beta = 0.178$, $t = 21.722$). Hedonic motivation, in turn, positively predicted both PE ($\beta = 0.389$, $t = 45.302$) and EE ($\beta = 0.194$, $t = 19.625$). Finally, EE had a positive effect on PE ($\beta = 0.112$, $t = 14.854$). Taken together, these paths suggest that social endorsement and perceived support help shape enjoyment, ease of use, and perceived usefulness, which are then associated with BI.

\subsubsection*{Total effects on BI}

To complement the direct-path results, the total effects on BI were examined (Table~\ref{tab:total_effects_bi}). The total effects combine direct and indirect pathways, and therefore provide a broader view of each variable's overall contribution to BI.

Social influence had by far the largest total effect on BI ($\beta_{\text{total}} = 0.825$, $t = 138.652$), reflecting both its direct association with intention and its indirect associations through other UTAUT2 constructs. Performance expectancy also had a substantial total effect ($\beta_{\text{total}} = 0.382$, $t = 25.615$), followed by HM ($\beta_{\text{total}} = 0.352$, $t = 26.354$), FC ($\beta_{\text{total}} = 0.277$, $t = 28.339$), and EE ($\beta_{\text{total}} = 0.097$, $t = 8.398$). The total-effect results show that HM, FC, and EE contribute more to BI than their direct effects alone suggest because part of their influence operates indirectly, especially through PE.

The background variables again had statistically significant, but small, total effects on BI--age ($\beta_{\text{total}} = –0.023$, $t = 7.099$), gender ($\beta_{\text{total}} = –0.030$, $t = 4.914$), and ADAS experience ($\beta_{\text{total}} = 0.028$, $t = 4.410$). These effects were modest relative to the total effects of the main UTAUT2 constructs.

\begin{table}[!ht]
\centering
\caption{Total effects on BI.}
\label{tab:total_effects_bi}
\begin{tabular}{lcc}
\hline
\textbf{Path} & \boldmath$\beta_{\text{total}}$ & \textbf{\emph{t}} \\
\hline
AGE $\rightarrow$ BI        & –0.023$^{*}$ &   7.099 \\
GENDER $\rightarrow$ BI     & –0.030$^{*}$ &   4.914 \\
EXPERIENCE $\rightarrow$ BI &  0.028$^{*}$ &   4.410 \\
EE $\rightarrow$ BI         &  0.097$^{*}$ &   8.398 \\
FC $\rightarrow$ BI         &  0.277$^{*}$ &  28.339 \\
HM $\rightarrow$ BI         &  0.352$^{*}$ &  26.354 \\
PE $\rightarrow$ BI         &  0.382$^{*}$ &  25.615 \\
SI $\rightarrow$ BI         &  0.825$^{*}$ & 138.652 \\
\hline
\end{tabular}
\begin{flushleft}
\footnotesize
\textit{Notes:} $\beta_{\text{total}}$ represents the standardized total effect on BI. $^{*}p < 0.001$.
\end{flushleft}
\end{table}

\subsubsection*{Moderating effects of age, gender, and ADAS experience}

Age, gender, and previous ADAS experience were tested as moderators of the relationships between the five UTAUT2 predictors and BI using interaction terms. The results show a selective moderation pattern (Table~\ref{tab:moderation}). Age moderated four of the five predictor--BI relationships. Specifically, age negatively moderated the FC $\rightarrow$ BI path ($\beta_{\mathrm{direct}} = –0.020$, $t = 3.323$, $p = 0.001$) and the SI $\rightarrow$ BI path ($\beta_{\mathrm{direct}} = –0.022$, $t = 3.142$, $p = 0.002$), indicating that these positive associations were weaker in older age categories. By contrast, age positively moderated the HM $\rightarrow$ BI path ($\beta_{\mathrm{direct}} = 0.027$, $t = 3.705$, $p < 0.001$) and the PE $\rightarrow$ BI path ($\beta_{\mathrm{direct}} = 0.017$, $t = 2.194$, $p = 0.028$), indicating that these positive associations were stronger in older age categories. The AGE $\times$ EE interaction was not statistically significant.

No statistically significant gender moderation effects were found, and previous ADAS experience also did not significantly moderate any of the five predictor--BI relationships. Thus, although gender and ADAS experience had small direct associations with BI, they did not significantly change the strength of the relationships between the UTAUT2 predictors and BI. Overall, the moderation results indicate that the central acceptance mechanism was broadly similar across gender and ADAS-experience groups, with modest age-related variation in selected paths.

\begin{table}[!ht]
\centering
\caption{Moderating effects on BI.}
\label{tab:moderation}

\begin{tabular}{lcc}
\hline
\textbf{Path} & \boldmath$\beta_{\mathrm{direct}}$ & \boldmath$t$ \\
\hline
AGE $\times$ EE $\rightarrow$ BI & 0.011 & 1.935 \\
AGE $\times$ FC $\rightarrow$ BI & –0.020** & 3.323 \\
AGE $\times$ HM $\rightarrow$ BI & 0.027*** & 3.705 \\
AGE $\times$ PE $\rightarrow$ BI & 0.017* & 2.194 \\
AGE $\times$ SI $\rightarrow$ BI & –0.022** & 3.142 \\
\hline
GENDER $\times$ EE $\rightarrow$ BI & 0.015 & 1.183 \\
GENDER $\times$ FC $\rightarrow$ BI & –0.003 & 0.229 \\
GENDER $\times$ HM $\rightarrow$ BI & 0.014 & 0.885 \\
GENDER $\times$ PE $\rightarrow$ BI & –0.019 & 1.173 \\
GENDER $\times$ SI $\rightarrow$ BI & 0.015 & 1.092 \\
\hline
EXPERIENCE $\times$ EE $\rightarrow$ BI & 0.017 & 1.376 \\
EXPERIENCE $\times$ FC $\rightarrow$ BI & 0.008 & 0.586 \\
EXPERIENCE $\times$ HM $\rightarrow$ BI & –0.005 & 0.354 \\
EXPERIENCE $\times$ PE $\rightarrow$ BI & –0.023 & 1.355 \\
EXPERIENCE $\times$ SI $\rightarrow$ BI & –0.014 & 1.027 \\
\hline
\end{tabular}

\begin{flushleft}
\footnotesize
\textit{Notes:} $\beta_{\mathrm{direct}}$ represents the standardized interaction effect. *$p < 0.05$, **$p < 0.01$, ***$p < 0.001$. For the interaction terms, the direct and total effects are identical because the moderation paths were specified only as direct paths to BI, and no indirect pathways from the interaction terms were included in the model.
\end{flushleft}

\end{table}

\subsection*{Predictive relevance and predictive performance}

A prediction-oriented assessment was conducted using PLSpredict and the CVPAT. Table~\ref{tab:plspredict_summary} summarizes the PLSpredict results for the endogenous construct indicators. All $Q^2_{\mathrm{predict}}$ values were positive, ranging from 0.239 to 0.554, indicating that the PLS-SEM model had predictive relevance relative to the naive mean-based prediction. The PLS-SEM model also produced lower RMSE and MAE values than the IA benchmark for all 19 indicators. However, it did not produce lower RMSE and MAE values than the LM benchmark for any indicator. Thus, PLSpredict showed predictive relevance relative to the IA benchmark, but not superior performance relative to the more conservative LM benchmark.

\begin{table}[!ht]
\begin{adjustwidth}{-2.25in}{0in}
\centering
\caption{PLSpredict summary for endogenous construct indicators}
\label{tab:plspredict_summary}

\begin{tabular}{lccccc}
\hline
\textbf{Construct} & \textbf{Indicators} & \textbf{$Q^2_{\mathrm{predict}}$ range} & \textbf{Mean $Q^2_{\mathrm{predict}}$} & \textbf{PLS $<$ IA} & \textbf{PLS $<$ LM} \\
\hline
BI & 5 & 0.375--0.554 & 0.477 & 5/5 & 0/5 \\
EE & 3 & 0.251--0.303 & 0.279 & 3/3 & 0/3 \\
FC & 4 & 0.239--0.297 & 0.270 & 4/4 & 0/4 \\
HM & 3 & 0.451--0.477 & 0.463 & 3/3 & 0/3 \\
PE & 4 & 0.454--0.484 & 0.472 & 4/4 & 0/4 \\
\hline
\end{tabular}

\begin{flushleft}
\textit{Notes:} IA, indicator average benchmark; LM, linear-model benchmark. The PLS $<$ IA and PLS $<$ LM columns report the number of indicators for which both the PLS-SEM RMSE and MAE values were lower than the corresponding benchmark values.
\end{flushleft}

\end{adjustwidth}
\end{table}

The CVPAT results supported the same interpretation (Table~\ref{tab:cvpat}). Compared with the IA benchmark, the ALDs were negative and statistically significant for all endogenous constructs. The overall PLS-SEM loss was 0.676, compared with an IA loss of 1.148, producing an ALD of –0.472 ($t = 60.566$, $p < 0.001$). This indicates that the model predicted significantly better than the naive IA benchmark.

By contrast, the CVPAT comparison with the LM benchmark showed positive and statistically significant ALDs for all endogenous constructs. The overall PLS-SEM loss was 0.676, whereas the LM loss was 0.649, producing an ALD of 0.027 ($t = 19.228$, $p < 0.001$). Because positive loss differences indicate higher prediction loss for PLS-SEM than for the LM benchmark, the model did not demonstrate superior predictive ability relative to the stricter LM benchmark.

\begin{table}[!ht]
\centering
\caption{CVPAT comparison of PLS-SEM with IA and LM benchmarks.}
\label{tab:cvpat}

\begin{tabular}{lcccccc}
\hline
\textbf{Construct} & \multicolumn{3}{c}{\textbf{PLS-SEM vs. IA}} & \multicolumn{3}{c}{\textbf{PLS-SEM vs. LM}} \\
\cline{2-4} \cline{5-7}
 & \textbf{ALD} & \boldmath$t$ & \boldmath$p$ & \textbf{ALD} & \boldmath$t$ & \boldmath$p$ \\
\hline
FC      & –0.272 & 39.374 & $<0.001$ & 0.028 & 15.138 & $<0.001$ \\
HM      & –0.560 & 51.986 & $<0.001$ & 0.039 & 16.380 & $<0.001$ \\
EE      & –0.258 & 36.606 & $<0.001$ & 0.018 & 12.336 & $<0.001$ \\
PE      & –0.551 & 54.863 & $<0.001$ & 0.028 & 14.169 & $<0.001$ \\
BI      & –0.644 & 64.389 & $<0.001$ & 0.024 & 15.564 & $<0.001$ \\
Overall & –0.472 & 60.566 & $<0.001$ & 0.027 & 19.228 & $<0.001$ \\
\hline
\end{tabular}

\begin{flushleft}
\footnotesize
\textit{Notes:} ALD, average loss difference; CVPAT, cross-validated predictive ability test. Negative ALD values indicate a lower average prediction loss for PLS-SEM than for the benchmark, whereas positive ALD values indicate a higher average prediction loss for PLS-SEM than for the benchmark. P-values reported as 0.000 in SmartPLS are presented as $p < 0.001$.
\end{flushleft}

\end{table}

Overall, the prediction-oriented results indicate that the model has predictive relevance and significantly outperformed the naive IA benchmark. However, it did not outperform the more conservative LM benchmark. The model's out-of-sample predictive power should therefore be interpreted as limited when assessed against the LM benchmark.

\section*{Discussion}

An extended UTAUT2 model was applied to a large pooled sample of respondents from 17 countries to examine the public acceptance of conditionally automated (SAE Level~3) cars. Using harmonized items from the L3Pilot questionnaire and PLS-SEM, the analysis extended previous UTAUT2-based work on Level~3 automation by estimating a single pooled structural model across the full first-phase multinational dataset. Overall, the measurement model showed good indicator reliability, internal consistency, and convergent validity. The structural model also showed strong explanatory power, accounting for 82.7\% of the variance in BI. However, discriminant validity was only partially supported because several theoretically related UTAUT2 constructs showed high HTMT values, even though all HTMT confidence intervals remained below 1.00 and all retained indicators loaded highest on their intended constructs.

\subsection*{Main drivers of BI}

In line with the UTAUT2, all five attitudinal constructs had positive and statistically significant direct effects on BI to use Level~3 cars. The strongest direct predictor was PE ($\beta = 0.382$), followed by SI ($\beta = 0.331$) and HM ($\beta = 0.185$). Effort expectancy ($\beta = 0.054$) and FC ($\beta = 0.028$) were also significant, but their direct effects were much smaller.

This pattern partly replicates previous UTAUT2-based research on automated vehicle acceptance. Nordhoff et al.\cite{Nordhoff2020} found that HM, SI, and PE were the three most important predictors of BI in a large European sample. In the present pooled 17-country model, PE and SI had larger direct effects than HM. This suggests that, across the broader multinational sample analyzed here, perceived usefulness and social endorsement were especially important drivers of intention to use conditionally automated cars, while anticipated enjoyment also made a meaningful contribution.

The comparatively small direct effects of EE and FC indicate that ease of use and perceived support matter, but they are not the primary direct levers of acceptance once perceived benefits, SI, and enjoyment are taken into account. Conceptually, the results suggest that the acceptance of Level~3 automation is driven less by the question of whether respondents think they can operate the technology, and more by whether they perceive it as useful, socially supported, and enjoyable.

The high explanatory power for BI ($R^2 = 0.827$) indicates that the reduced UTAUT2 model, even without price value and habit, captured a substantial share of the systematic variance in intention to use Level~3 cars. This supports the usefulness of the UTAUT2 for modeling conditional-automation acceptance in a large multinational dataset, while also showing that the strongest predictors are concentrated around usefulness, social endorsement, and enjoyment. Because a pooled structural model was reported rather than a multi-group analysis, these findings should be viewed as overall relationships in the combined sample, not as evidence of country- or region-specific differences.

\subsection*{Relation to sustainable and connected-vehicle adoption research}

Our findings also align with recent research that framed AV adoption as part of a broader sustainability and connected-mobility transition. In our pooled model, PE was the strongest direct predictor of BI, indicating that the respondents were more willing to use Level~3 cars when they perceived clear functional benefits. This is consistent with sustainability-oriented AV research showing that perceived economic benefits, environmental benefits, and environmental concerns are important drivers of green behavior and sustainable adoption intentions\cite{Arpaci2024,Arpaci2026}. Although the L3Pilot data do not directly measure environmental benefits or economic value, the strong role of PE suggests that users' perceived benefits are central to their acceptance, whether those benefits are framed as comfort, safety, convenience, cost savings, or sustainability.

The large direct and total effects of SI also resonate with recent evidence on authority support, subjective norms, and behavioral reasoning in sustainable mobility adoption. Arpaci et al.\cite{Arpaci2025,Arpaci2026} showed that perceived authority support contributes to sustainability-oriented BIs toward AV use, while the V2X adoption study showed that environmental values shape reasons for and against adoption, which then influence attitudes and BI. In our study, SI was not limited to formal authority support, rather capturing broader perceived social endorsement from important others. Nevertheless, the finding that SI had the greatest total effect on BI suggests that the acceptance of Level~3 automation may be socially constructed through interpersonal approval, public narratives, institutional confidence, and perceived legitimacy.

The relatively small direct effects of EE and FC should also be interpreted in light of the sustainability and V2X literature. The V2X adoption study highlighted technology-specific reasons for and against adoption, including perceived advantages, concerns, and barriers\cite{Arpaci2025}. Similarly, the innovation resistance theory study on AVs emphasized cost barriers and security and privacy concerns as negative predictors of green behavior\cite{Arpaci2024}. The L3Pilot UTAUT2 model did not include these resistance constructs directly. Therefore, the modest direct effects of EE and FC should not be read as evidence that support conditions, cost, cybersecurity, privacy, or infrastructure are unimportant. Rather, among the variables measured here, ease of use and support played smaller direct roles than usefulness, SI, and enjoyment, while also contributing indirectly through other constructs.

Together, these comparisons suggest that the UTAUT2 captures the core acceptance mechanism for Level~3 cars, while recent studies based on the theory of planned behavior, innovation resistance theory, and behavioral reasoning theory have identified additional sustainability and resistance mechanisms that should be incorporated into future models. A more comprehensive model of sustainable AV adoption would combine UTAUT2 constructs with environmental values, perceived environmental benefits, economic benefits, authority support, cost barriers, cybersecurity and privacy concerns, and reasons for and against connected-vehicle technologies, such as V2X.

\subsection*{Interrelations among UTAUT2 constructs}

The structural model also revealed strong interrelations among the UTAUT2 predictor constructs. Social influence acted as a central upstream driver. It had a strong positive effect on FC ($\beta = 0.653$), HM ($\beta = 0.519$), and PE ($\beta = 0.307$), as well as a smaller, albeit significant, effect on EE ($\beta = 0.127$). These findings indicate that social endorsement affects BI not only directly, but also indirectly, by shaping perceptions of support, enjoyment, ease of use, and usefulness.

Facilitating conditions also played an important upstream role, strongly predicting EE ($\beta = 0.538$) and having additional positive effects on HM ($\beta = 0.364$) and PE ($\beta = 0.178$). Hedonic motivation, in turn, predicted both PE ($\beta = 0.389$) and EE ($\beta = 0.194$), while EE predicted PE ($\beta = 0.112$). These results suggest a layered acceptance process in which social endorsement and perceived enabling conditions shape enjoyment and ease of use, which then contribute to perceived usefulness and BI.

The total-effect results reinforce this interpretation, with SI having by far the largest total effect on BI ($\beta_{\text{total}} = 0.825$), showing that much of its importance comes from indirect pathways through other constructs. Performance expectancy also had a substantial total effect on BI ($\beta_{\text{total}} = 0.382$), followed by HM ($\beta_{\text{total}} = 0.352$), FC ($\beta_{\text{total}} = 0.277$), and EE ($\beta_{\text{total}} = 0.097$). Thus, although PE was the strongest direct predictor of BI, SI was the dominant overall driver when indirect effects were considered.

\subsection*{Measurement overlap and construct interpretation}

The measurement model results support reliability and convergent validity, but they also show that several UTAUT2 constructs are highly related in this application. The HTMT values exceeded the stricter 0.90 threshold for PE--BI, SI--BI, EE--FC, HM--BI, and PE--HM. At the same time, all bootstrapped HTMT confidence intervals remained below 1.00, and all indicators loaded highest on their intended constructs. Discriminant validity was therefore not absent, but it was only partially supported under stricter HTMT-based criteria.

This pattern is substantively important. The respondents may not have experienced PE, SI, HM, and BI as fully separate psychological dimensions when evaluating an unfamiliar technology, such as Level~3 automation. Instead, these constructs may partly reflect a broader favorable orientation toward conditionally automated cars. This does not invalidate the structural model, but it does mean that small differences among the PE, SI, and HM coefficients should be interpreted cautiously because these predictors are strongly correlated.

Future research could address this issue in two ways. First, studies could test higher- or second-order structures---for example, by modeling a broader perceived-benefit or positive-attitude factor that includes performance-related and affective beliefs. Second, future surveys could use more context-specific items that distinguish between the distinctive features of Level~3 automation, such as take-over requests, operational design domains, perceived fallback responsibility, and trust in the handover process. Such refinements may help reduce overlap among UTAUT2 constructs while preserving their explanatory value.

\subsection*{Demographic and ADAS-experience effects}

Age, gender, and ADAS-use experience had statistically significant, but very small, direct effects on BI. Age had a small negative effect ($\beta = –0.023$), indicating that respondents in older age categories reported slightly lower intention to use Level~3 cars. Gender also had a small negative effect under the coding used in the analysis ($\beta = –0.030$), meaning that respondents coded as female reported slightly lower BI than respondents coded as male, although the magnitude of this association was very small. Advanced driver assistance systems-use experience had a small positive effect ($\beta = 0.028$), suggesting that those respondents who had experience using a greater number of driver assistance systems were somewhat more open to Level~3 automation. These background-variable effects were much smaller than the effects of the main UTAUT2 constructs, indicating that demographic and experience-related differences are secondary to perceived usefulness, social endorsement, and enjoyment.

The moderation results added nuance to this pattern. Age significantly moderated four of the five predictor--intention relationships, although the interaction coefficients were small. Age weakened the positive associations of FC ($\beta = –0.020$) and SI ($\beta = –0.022$) with BI, but strengthened the positive associations of HM ($\beta = 0.027$) and PE ($\beta = 0.017$) with BI. The age $\times$ EE interaction was not statistically significant.

By contrast, neither gender nor ADAS-use experience significantly moderated any of the five UTAUT2 predictor--intention relationships. Thus, the central acceptance mechanism was broadly similar across gender and ADAS-experience groups, with only modest age-related variation in the strength of selected predictors. In practical terms, age, gender, and ADAS-use experience may help identify groups requiring tailored communication or additional support, but the main acceptance levers remain perceived usefulness, social endorsement, and enjoyment.

The ADAS-use experience count had a small positive effect ($\beta = 0.028$), suggesting that those respondents who had experienced using a greater number of driver assistance systems were somewhat more open to Level~3 automation. However, this effect was much smaller than the effects of the main UTAUT2 constructs. Previous ADAS-use breadth therefore appears to be a secondary background factor rather than a central determinant of acceptance.

\subsection*{Implications for design, policy, and communication}

Our findings suggest several practical levers for increasing the informed public acceptance of Level~3 automation. First, the strong direct role of PE indicates that communication should focus on concrete and credible benefits. Potential users need to understand how conditional automation may improve driving or travel---for example, by making such travel feel more useful, comfortable, convenient, and effective under clearly defined operating conditions. Such communication should avoid overpromising, especially because Level~3 systems still require the human driver to resume control when requested.

Second, the large direct and total effects of SI show that social endorsement is central to acceptance. Public acceptance may therefore be shaped not only by technical information from manufacturers or regulators, but also by trusted interpersonal sources, visible early adopters, media narratives, and demonstrations. Because SI also strongly predicted FC, HM, and PE, social endorsement appears to shape how people evaluate the broader feasibility, enjoyment, and usefulness of Level~3 cars.

Third, HM's meaningful direct effect on BI and its strong effects on PE and EE highlight the importance of user experience. Level~3 systems should not only function safely and reliably, but should also feel comfortable, intuitive, and confidence-building. Enjoyment should be understood broadly here, not merely as entertainment, but as a positive user experience that supports trust, perceived usefulness, and ease of interaction. At the same time, designers must ensure that enjoyable non-driving-related activities do not undermine driver readiness to respond to take-over requests.

Fourth, FC and EE had smaller direct effects on BI, but they had important indirect roles, with FC strongly predicting EE and also contributing to HM and PE, while EE contributed to PE. This suggests that training, support, clear instructions, transparent system boundaries, and accessible help channels can indirectly improve acceptance by making Level~3 cars feel easier to use, more enjoyable, and more useful. Acceptance is therefore not only a matter of vehicle design, but also of the surrounding ecosystem, including infrastructure readiness, legal clarity, consumer education, and after-sales support.

Finally, the recent sustainability and V2X adoption literature suggests that communication strategies should not focus only on convenience or technological novelty. Messages about Level~3 automation should also explain any potential environmental and economic benefits, where these are empirically justified, clarify the role of public authorities and regulatory safeguards, and address concerns about cost, cybersecurity, privacy, and connected-vehicle infrastructure\cite{Arpaci2024,Arpaci2025,Arpaci2026}. For conditionally automated cars, this communication must remain precise---Level~3 systems are not fully driverless, and users must understand the operational design domains, take-over requests, and fallback responsibilities. Sustainable adoption therefore depends on both positive value propositions and the credible management of technology-specific risks.

The prediction-oriented results qualified the model's strong explanatory performance. Although the model explained a large share of variance in BI, and all the $Q^2_{\mathrm{predict}}$ values were positive, the PLSpredict and CVPAT comparisons showed that the model's predictive advantage was mainly relative to the naive IA benchmark. The LM benchmark produced lower prediction loss than the PLS-SEM model. Thus, the findings provide strong explanatory evidence for the UTAUT2 acceptance mechanism and evidence of predictive relevance, but they do not show that the PLS-SEM specification is predictively superior to a simpler linear benchmark.

The supplementary global fit results were also broadly consistent with this interpretation, with the estimated-model SRMR being 0.048, indicating good approximate fit, although the NFI was slightly below the conventional 0.90 guideline.

\subsection*{Strengths, limitations, and directions for future research}

This study has several strengths. In using a large harmonized multinational dataset of licensed drivers from 17 countries, it was able to examine the public acceptance of SAE Level~3 cars across a broader international sample than many previous single-country or region-specific studies. The study also focused specifically on conditional automation rather than aggregating multiple automation levels, applied an established UTAUT2-based measurement framework, evaluated both the measurement and structural models, and distinguished direct from total effects to clarify both proximal and indirect acceptance mechanisms. In addition, the use of PLSpredict and the CVPAT provided a prediction-oriented check on the explanatory model, rather than relying only on in-sample path estimates.

These strengths should be considered alongside several limitations when assessing the findings. First, the study was reliant on cross-sectional survey data, which limited our ability to draw causal conclusions about the relationships among the constructs. Although the results showed significant associations between PE, EE, SI, HM, FC, ADAS-use experience, and intention to use SAE Level~3 conditionally automated cars, the design did not establish whether these factors caused changes in acceptance. Public attitudes toward Level~3 automation may also evolve over time as people gain more exposure to ADAS, automated driving functions, and conditional automation. Future research should therefore use longitudinal designs to examine how acceptance changes as Level~3 systems become more familiar and more widely available.

Second, the outcome measured in this study reflects self-reported BI, rather than the actual adoption or real-world use of SAE Level~3 conditionally automated cars. Intentions are useful for understanding public readiness, especially in contexts where Level~3 systems are not yet widely available, but they may not fully predict future behavior. Actual adoption may depend on additional factors, such as vehicle cost, perceived safety after direct exposure, insurance conditions, regulatory approval, infrastructure readiness, and the availability of suitable Level~3 functions. Future studies should complement intention-based survey research with behavioral data, field experiments, pilot studies, or naturalistic exposure to conditionally automated vehicles.

Third, the study used quota sampling through online panels. While this approach made it possible to collect a large international sample of more than 18{,}000 respondents across 17 countries, online panel recruitment can introduce selection bias. Individuals who participate in online panels may differ from the broader driving population in terms of digital literacy, technology interest, socioeconomic status, education, and access to internet-based survey platforms. In addition, although quotas can improve sample balance on selected demographic characteristics, they do not guarantee full national representativeness. Therefore, the results should be viewed as evidence from a large cross-national sample rather than as population-level estimates for each country. Future research could strengthen its generalizability by using probability-based sampling, post-stratification weights, or comparisons with national demographic benchmarks.

Fourth, our treatment of missing data represented a methodological limitation. Responses coded as ``I prefer not to respond'' were recoded as missing values, and remaining missing values in the model variables were handled in SmartPLS~4 using mean replacement. Although this approach allowed the full retained analytic sample to be used for model estimation, mean replacement can attenuate variance, reduce covariance among indicators, and understate uncertainty if the missingness is systematic. Future studies should examine missing-data mechanisms more explicitly and consider alternative approaches, such as multiple imputation, full information maximum likelihood, or sensitivity analyses comparing mean replacement with complete-case and imputed results.

Fifth, while the study included respondents from 17 countries, cross-national interpretation remains challenging. Because our analysis estimated a pooled structural model and did not report country- or region-level multi-group results, the findings should not be viewed as evidence that the model operated identically across all national contexts. Differences in acceptance of Level~3 conditionally automated cars may reflect not only individual-level beliefs, but also broader country-specific conditions, including road infrastructure, traffic complexity, urban density, public transport systems, regulatory frameworks, safety culture, media coverage, and previous exposure to vehicle automation. These contextual factors were not directly measured in our survey and therefore could not be included as explanatory variables in the main model. Future research should integrate individual-level survey data with country-level indicators, such as infrastructure quality, road safety statistics, congestion levels, AV policy readiness, and national innovation capacity, to better explain why acceptance varies across countries.

Sixth, measurement comparability across countries is an important issue in international survey research. Because constructs such as PE, SI, trust, perceived ease of use, and HM may be interpreted differently across cultural and linguistic contexts, future cross-national studies should continue to examine measurement invariance across countries or regions. Establishing configural, metric, and scalar invariance is important for determining whether observed differences reflect true differences in acceptance, or differences in how respondents interpret the measurement items.

Seventh, the study did not fully capture the experience of drivers who had directly used SAE Level~3 conditionally automated vehicles. Although ADAS experience can provide useful insights into the respondents' familiarity with vehicle automation, ADAS use is not equivalent to direct experience with conditional automation. Acceptance may change substantially when drivers experience the capabilities, limitations, handover requirements, and safety implications of SAE Level~3 systems under real-world conditions. Future research should interrogate users with direct exposure to SAE Level~3 vehicles and compare them with drivers who have experience only with lower-level assistance systems.

Eighth, the study was based on self-reported survey measures, which may have been affected by common method bias, social desirability bias, and the respondents' subjective understanding of automated-vehicle technology. Some respondents may have had limited technical knowledge of the differences between ADAS, partial automation, conditional automation, and fully driverless automation. As a result, their responses may reflect general attitudes toward innovation or vehicle technology rather than being precise evaluations of SAE Level~3 systems. Future studies could reduce this limitation by providing standardized descriptions, visual scenarios, experimental vignettes, or demonstrations before measuring acceptance.

Ninth, the L3Pilot dataset does not contain several constructs that have become increasingly important in recent research on sustainable AV and V2X adoption. In particular, the study could not directly model environmental values, green behavior, environmental sustainability outcomes, perceived authority support, perceived economic benefits, cost barriers, cybersecurity concerns, privacy concerns, or explicit reasons for or against V2X adoption\cite{Arpaci2024,Arpaci2025,Arpaci2026}. As a result, while our study explained the general BI to use Level~3 cars, it could not determine whether such intention translated into environmentally sustainable behavior or acceptance of connected-vehicle functions. Future research should integrate the UTAUT2 with the theory of planned behavior, innovation resistance theory, and behavioral reasoning theory to examine how perceived usefulness, SI, enjoyment, environmental values, economic value, institutional support, and resistance factors jointly shape sustainable AV adoption.

Finally, automated-vehicle technology, regulation, and public debate are developing rapidly. Public acceptance of Level~3 systems may be influenced by recent news, safety incidents, commercial deployments, government policy changes, and broader social narratives about artificial intelligence and automation. As a result, the findings should be viewed as reflecting attitudes during the period in which the data were collected. Continued research is needed to track how public acceptance changes as conditionally automated vehicles move from experimental and limited deployment settings toward broader commercial availability.

Overall, these limitations did not undermine the value of the study, but they indicate that the findings should be viewed with appropriate caution. While the large 17-country sample provides important evidence on pooled patterns of Level~3 conditionally automated car acceptance, future research can build on this work by using longitudinal designs, nationally representative samples, direct behavioral measures, country-level contextual data, multi-group analyses, and studies involving drivers with real-world experience of conditional automation. The prediction-oriented assessment produced a more cautious conclusion, the model showing predictive relevance and significantly outperforming the IA benchmark, but it did not outperform the LM benchmark. Therefore, while the study has provided strong evidence for the explanatory usefulness of the UTAUT2-based model, its out-of-sample predictive superiority over a simpler linear benchmark remains limited.

\subsection*{Conclusion}

The results showed that a UTAUT2-based model provides a strong explanation of BI to use SAE Level~3 conditionally automated cars in a large pooled multinational sample. Perceived usefulness, SI, and HM were the dominant direct predictors of BI, while EE and FC played smaller direct roles. When indirect pathways were considered, SI emerged as the strongest overall driver of intention. Age, gender, and ADAS-use experience were statistically significant, but small, direct predictors. Age also produced several small moderation effects, whereas gender and ADAS-use experience did not significantly moderate the UTAUT2 predictor--intention relationships. Overall, demographic and experience-related differences were secondary to the main attitudinal drivers.

These findings suggest that public acceptance of conditional automation depends primarily on whether people view the technology as useful, socially supported, enjoyable, and embedded in an enabling use environment. For design, policy, and communication, the results point to the importance of demonstrating concrete benefits, building social confidence, supporting positive user experiences, and ensuring that users understand the resources, system boundaries, and responsibilities involved in using Level~3 automation safely and effectively.

The findings should nevertheless be viewed with appropriate caution. A pooled cross-sectional model was used and therefore did not establish causal effects or country-specific acceptance mechanisms. In addition, several UTAUT2 constructs were highly related, and the model showed stronger explanatory performance than predictive superiority over a LM benchmark. Future research should therefore combine the UTAUT2 with additional sustainability, resistance, trust, cost, cybersecurity, privacy, and policy-related constructs, especially as Level~3 systems move from pilot settings toward broader market availability.

\nolinenumbers

%
%
%

\section*{Data availability}
The data analyzed in this study are publicly available. The L3Pilot Global User Acceptance Survey, First Phase Data, which is hosted on Zenodo was used, and can be accessed at \url{https://zenodo.org/records/8389544}. The dataset contains fully anonymized survey responses from licensed drivers across 17 countries and is distributed under an open license with no restrictions on access or reuse.

\section*{Author contributions (CRediT)}
Conceptualization: A.S.\\
Methodology: A.S.\\
Formal analysis: A.S.\\
Writing -- original draft: A.S.\\
Writing -- review \& editing: A.S.

\section*{Supporting information}

\begin{table}[!ht]
\centering
\caption*{S1 Table. Initial outer loadings from the first measurement model.}
\begin{tabular}{llc}
\hline
\textbf{Construct} & \textbf{Indicator} & \textbf{Initial outer loading} \\
\hline
\textbf{Behavioral Intention (BI)} & BI1 & 0.880 \\
 & BI2 & 0.841 \\
 & BI3 & 0.729 \\
 & BI4 & 0.846 \\
 & BI5 & 0.819 \\
\textbf{Effort Expectancy (EE)} & EE1 & 0.852 \\
 & EE2 & 0.828 \\
 & EE3 & 0.861 \\
\textbf{Facilitating Conditions (FC)} & FC1 & 0.820 \\
 & FC2 & 0.803 \\
 & FC3 & 0.813 \\
 & FC4 & 0.747 \\
\textbf{Hedonic Motivation (HM)} & HM1 & 0.889 \\
 & HM2 & 0.905 \\
 & HM3 & 0.903 \\
\textbf{Performance Expectancy (PE)} & PE1 & 0.655 \\
 & PE2 & 0.858 \\
 & PE3 & 0.852 \\
 & PE4 & 0.868 \\
 & PE5 & 0.868 \\
\textbf{Social Influence (SI)} & SI1 & 0.861 \\
 & SI2 & 0.840 \\
 & SI3 & 0.859 \\
 & SI4 & 0.826 \\
\hline
\end{tabular}

\begin{flushleft}
\footnotesize
\textit{Notes}: Values are standardized outer loadings from the initial PLS-SEM measurement model before item removal. PE1 showed the weakest loading among the PE indicators and was subsequently removed from the final measurement model. Single-indicator observed variables and interaction terms had fixed outer loadings of 1.00 and are therefore not shown.
\end{flushleft}
\end{table}

\begin{table}[htbp]
\centering
\caption*{S2 Table. Outer loadings and indicator reliability after item removal.}
\begin{tabular}{llcc}
\hline
\textbf{Construct} & \textbf{Indicator} & \textbf{Outer Loading} & \textbf{$t$} \\
\hline
\textbf{Behavioral Intention (BI)} & BI1 & 0.880$^{*}$ & 397.754 \\
 & BI2 & 0.841$^{*}$ & 303.728 \\
 & BI3 & 0.729$^{*}$ & 151.454 \\
 & BI4 & 0.846$^{*}$ & 291.714 \\
 & BI5 & 0.818$^{*}$ & 267.542 \\
\textbf{Effort Expectancy (EE)} & EE1 & 0.852$^{*}$ & 263.464 \\
 & EE2 & 0.828$^{*}$ & 236.061 \\
 & EE3 & 0.861$^{*}$ & 318.666 \\
\textbf{Facilitating Conditions (FC)} & FC1 & 0.820$^{*}$ & 227.124 \\
 & FC2 & 0.803$^{*}$ & 213.327 \\
 & FC3 & 0.813$^{*}$ & 211.745 \\
 & FC4 & 0.747$^{*}$ & 157.676 \\
\textbf{Hedonic Motivation (HM)} & HM1 & 0.889$^{*}$ & 361.162 \\
 & HM2 & 0.905$^{*}$ & 459.203 \\
 & HM3 & 0.903$^{*}$ & 471.046 \\
\textbf{Performance Expectancy (PE)} & PE2 & 0.869$^{*}$ & 336.859 \\
 & PE3 & 0.863$^{*}$ & 317.110 \\
 & PE4 & 0.876$^{*}$ & 359.354 \\
 & PE5 & 0.881$^{*}$ & 377.275 \\
\textbf{Social Influence (SI)} & SI1 & 0.861$^{*}$ & 318.840 \\
 & SI2 & 0.840$^{*}$ & 254.585 \\
 & SI3 & 0.859$^{*}$ & 295.598 \\
 & SI4 & 0.826$^{*}$ & 291.757 \\
\hline
\end{tabular}

\begin{flushleft}
\footnotesize
\textit{Notes}: Values are standardized outer loadings from the final measurement model after removing PE1. $^{*}$ All outer loadings are statistically significant at $p < 0.001$ (two-tailed test).
\end{flushleft}
\end{table}

\begin{table*}[!ht]
\centering
\caption*{S3 Table. Summary of indicator cross-loadings.}
\label{tab:crossloadings}
\begin{tabular}{lcc}
\hline
\textbf{Item} & \textbf{Outer loading} & \textbf{Highest cross-loading} \\
\hline
BI1 & 0.880 & PE (0.757) \\
BI2 & 0.841 & PE (0.770) \\
BI3 & 0.729 & SI (0.613) \\
BI4 & 0.846 & PE (0.749) \\
BI5 & 0.818 & SI (0.714) \\
EE1 & 0.852 & FC (0.626) \\
EE2 & 0.828 & FC (0.618) \\
EE3 & 0.861 & FC (0.677) \\
FC1 & 0.820 & EE (0.666) \\
FC2 & 0.803 & PE (0.628) \\
FC3 & 0.813 & EE (0.647) \\
FC4 & 0.747 & PE (0.553) \\
HM1 & 0.889 & BI (0.720) \\
HM2 & 0.905 & PE (0.734) \\
HM3 & 0.903 & PE (0.771) \\
PE2 & 0.869 & BI (0.750) \\
PE3 & 0.863 & BI (0.739) \\
PE4 & 0.876 & BI (0.745) \\
PE5 & 0.881 & BI (0.771) \\
SI1 & 0.861 & BI (0.706) \\
SI2 & 0.840 & BI (0.655) \\
SI3 & 0.859 & BI (0.683) \\
SI4 & 0.826 & BI (0.768) \\
\hline
\end{tabular}

\begin{flushleft}
\footnotesize
\textit{Notes}: Outer loadings and cross-loadings were obtained from the SmartPLS measurement model. Each measurement item loads more strongly on its intended construct than on any other reflective construct, supporting indicator-level discriminant validity.
\end{flushleft}
\end{table*}

\end{document}